*Article*

# Flexibility of Fluorinated Graphene-Based Materials

**Irina Antonova [1,2,3,*], Nadezhda Nebogatikova [1,2], Nabila Zerrouki [4], Irina Kurkina [5]** 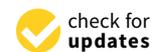 **and Artem Ivanov [1]**

[1] Rzhanov Institute of Semiconductor Physics SB RAS, Lavrentiev av. 13, 630090 Novosibirsk, Russia; nadonebo@gmail.com (N.N.); art.iv.il@mail.ru (A.I.)

[2] Physical Department, Novosibirsk State University, Pirogov str. 2, 630090 Novosibirsk, Russia

[3] Faculty of Radio Engineering and Electronics, Novosibirsk State Technical University, K. Marx str. 20, 630073 Novosibirsk, Russia

[4] Department of Chemistry, College of Science & Technology, The University of Bordeaux, 351 Cours de la Libération, 33405 Talence Cedex, France; nabila.zerrouki@hotmail.fr

[5] Institute of Physics and Technologies, Ammosov North-Eastern Federal University, Belinskii str. 58, 677000 Yakutsk, Russia; volkiraly@mail.ru

* Correspondence: antonova@isp.nsc.ru

**Abstract:** The resistivity of different films and structures containing fluorinated graphene (FG) flakes and chemical vapor deposition (CVD)-grown graphene of various fluorination degrees under tensile and compressive strains due to bending deformations was studied. Graphene and multilayer graphene films grown by means of the chemical vapor deposition (CVD) method were transferred onto the flexible substrate by laminating and were subjected to fluorination. They demonstrated a weak fluorination degree (F/C lower 20%). Compressive strains led to a strong (one-two orders of magnitude) decrease in the resistivity in both cases, which was most likely connected with the formation of additional conductive paths through fluorinated graphene. Tensile strain up to 3% caused by the bending of both types of CVD-grown FG led to a constant value of the resistivity or to an irreversible increase in the resistivity under repeated strain cycles. FG films created from the suspension of the fluorinated graphene with a fluorination degree of 20–25%, after the exclusion of design details of the used structures, demonstrated a stable resistivity at least up to 2–3% of tensile and compressive strain. The scale of resistance changes ΔR/R0 was found to be in the range of 14–28% with a different sign at the 10% tensile strain (bending radius 1 mm). In the case of the structures with the FG thin film printed on polyvinyl alcohol, a stable bipolar resistive switching was observed up to 6.5% of the tensile strain (bending radius was 2 mm). A further increase in strain (6.5–8%) leads to a decrease in ON/OFF current ratio from 5 down to 2 orders of magnitude. The current ratio decrease is connected with an increase under the tensile strain in distances between conductive agents (graphene islands and traps at the interface with polyvinyl alcohol) and thickness of fluorinated barriers within the active layer. The excellent performance of the crossbar memristor structures under tensile strain shows that the FG films and structures created from suspension are especially promising for flexible electronics.

## 1. Introduction

The rapid growth in the popularity of the Internet of things, particularly flexible, wearable and stretchable electronics, in the past decade has resulted in the strong progress in flexible electronics called flextronics [1]. It is expected that flextronics, in the near future, will take a significant place in our daily life, including health care [2–4], computation and memory [5], gadgets, touch screens and displays [6], energy storage and generation [7], electronic textile [8,9], human activities [1,10], and etc. Two-dimensional (2D) layered crystals are believed to be the most promising candidates for flextronic applications, owing to their well-known features including the ultimate thickness scalability down to the atomic level, the flexibility increasing with the decreasing layer thickness and high intrinsic strain limit [11]. In the family of 2D materials, this statement is attributed to the graphene, graphene oxide and transition metal dichalcogenides. However, it was found that, in striking contrast to graphene, the mechanical properties of insulating hexagonal boron nitride (h-BN) (dielectric crystals which provided the high carrier mobility in graphene) are not practically changed with reducing the BN thickness [12]. This results in the limitation of BN applications for flextronics [13,14]. So, nowadays, the problem of insulators for flextronics is still open. We have suggested that fluorinated graphene (FG) is one of the most promising insulators for flexible electronics [15,16]. Fluorinated graphene is often considered as a prospective material for applications in combination with graphene [17,18]. Generally, many 2D layered crystals, their derivatives and composites are believed to be prospective materials for flextronics, but the investigation of their mechanical properties remains incomplete or, for some materials, simply unknown.

The mechanical properties of graphene, the most investigated 2D material, have been widely discussed (see, for instance, [19]). Perfect graphene is robust and flexible. Chemical vapor deposition (CVD)-grown graphene was transferred onto the flexible substrate polydimethylsiloxane and showed no appreciable resistance change when applying a stretching deformation up to 12% [20]. According to the theoretical investigation, an asymmetric stretching changes the band structure and leads to the bandgap appearance: the stretching in the plane of C-C bonds leads to the maximum value of the bandgap being about 0.49 eV at $\varepsilon$ = 12.2% [21]. If one applied a stretching force perpendicularly to the C-C bonds, the maximum value of the bandgap will be 0.170 eV and will be reached with the deformation of 7.3% [22]. Along uniaxial strain, the band gaps appear at $\varepsilon$ = 23%.

In the case of the bending, the main effect is a change in the graphene resistivity. The graphene response on a tensile or compressive strain is expected to depend on structural features of the tested films (graphene or multilayer graphene CVD-grown films, or films created from the suspension). For instance, in [23,24], the optimal shape of graphene flakes in the composite film for their maximum flexibility was analyzed. The authors found that graphene flakes should have a few-layer shape. The Raman peaks of monolayer, bilayer and multilayer graphene flakes consist of multiple components that exhibit different shift behaviors—the rate of G and 2G band shift is lower for a few-layer specimen than that of a monolayer one, indicating a relatively poor stress transfer between the graphene layers. This type of flake architecture provides the base for the optimization of the strain transfer efficiency of the graphene-based composites due to the prevention of interlayer sliding.

Fluorographene is known to have a relatively low Young's modulus of E = 100 ± 30 N/m, and intrinsic strength $\sigma$ = 15 N/m [25]. For comparison, graphene has E = 340 ± 50 N/m and $\sigma$ = 42 ± 4 N/m, respectively [26]. The flexibility of partially fluorinated graphene (FG) was analyzed in our previous study [16], and it was demonstrated that, for thin insulating films (~20 nm) created from a suspension with the use of injecting printing, the unchanged leakage current (~ $10^{-7}$ A/cm$^2$) was found up to the strain of 10%.

In the present study, the flexibility of partially fluorinated graphene (CVD-grown graphene and multilayer graphene (MG) or films created from the FG suspension) with the pyramid-like shape was compared with that for the films created from FG the suspension. Excellent mechanical properties of the films created from fluorinated graphene with the relatively low fluorinated degree were found. The formation of FG films on polyvinyl acid films resulted in the creation of the structures with resistive switches. The operation of these structures, in the case of tensile bending, demonstrates the stable resistive switches down to the radius of about 2 mm (strain~6.5%) with a current decrease in the open state.

## 2. Experimental Methods

### 2.1. Fluorination of CVD Grown Graphene and Multilayer Graphene

Graphene and multilayer graphene were CVD grown on the copper foil at the temperature of 950–1050 °C. The domain size in these polycrystalline films was ~1–5 m. The reference sheet resistance measured by the four-probe heads for graphene and MG were 1–2 kOhn/sq and 4–10 kOhn/sq, respectively. The fluorination process of graphene proceeds by means of treatment in a weak (~3–7%) solution of hydrofluoric acid. The specialties for the chosen fluorination conditions are given in detail in Ref. [25,26] and in the references given in these papers. Fluorination time was chosen depending on the film thickness (increase in the multilayer graphene thickness led to the increase in the fluorination time). A strong increase in the film resistivity corresponds to the fluorination degree of F/C ~ 25%, which was determined from X-ray photoelectron spectroscopy (XPS) [27].

### 2.2. Preparation of Fluorinated Graphene Suspensions

An important starting step in the preparation of fluorinated graphene suspensions is the synthesis of graphene suspension. We used the liquid exfoliation of natural graphite by means of a dispergator combined with ultrasonic processing and centrifugation. Then, the suspension filtration aimed at eliminating large-sized flakes was performed. The track membranes used for separation have pore sizes of 1.2 μm.

The fluorination process of graphene suspension also proceeds in a weak (~3–7%) solution of hydrofluoric acid. During the fluorination process, due to additional exfoliation, the thickness of fluorinated graphene flakes strongly decreases down to 1–5 monolayers. The fluorination degree was approximately estimated from a suspension color and then, more exactly, was determined by means of the XPS measurements. After the fluorination of the suspension, the hydrofluoric acid residue was removed from the solution by means of displacement in water. The films from the FG suspension were created on Si or other desirable substrates by means of drops with the same volume (0.1 mL) or 2D printing for testing.

The polyvinyl alcohol films were prepared by applying a 1% PVA solution onto the surface of a silicon wafer in an implemented spinning process. The molecular mass of the used PVA was 13,000.

The films were printed on a DMP-2831 Dimatix FUJIFILM jet printer (Fujifilm, Lebanon, PA, USA). After printing a layer, the film was subjected to a drying procedure at 60 °C to exclude possible mixing effects.

A Solver PRO NT-MDT scanning microscope (NT-MDT, Moscow, Russia) was used for obtaining the atomic force microscopy (AFM) images of suspension flakes and film surfaces, and for determining the film thicknesses. The measurements were carried out in both contact and semi-contact mode. Silicon cantilever tips (HA_C/Au) with the typical resonance frequency values of 19 and 37 kHz and force constants of 0.26 and 0.65 N/m, respectively, and a tip radius of less than 10 nm were used. The differentiation of the regions, according to their chemical composition, was based on the images recorded in the lateral-force (friction-force) mode. The Raman spectra were recorded at room temperature; the excitation wavelength was 514.5 nm (2.41-eV argon ion laser). In order to avoid the heating of the sample with laser radiation, the laser beam power was decreased to 2–3 mW. The scanning electron microscopy (SEM) images were obtained using a JEOL JSM-7800F scanning electron microscope (Jeol, Tokyo, Japan) with the energy of primary electrons equal to 2 keV. JEOL JSM 7800F was equipped with a super-hybrid objective lens and the "Gentle Beam" system, which allows for obtaining high-resolution images at low accelerating voltages. The use of such systems makes it possible to obtain the film images without the deposition of electrically conductive coatings.

The sheet resistance of the obtained films was studied using the four-probe JANDEL equipment and HM21 Test Unit at room temperature (Jandel Engineering Limited, Limslade, UK). The capacitance-voltage (C-V) and current-voltage (I-V) characteristics of the fabricated structures were measured using a precision LCR meter E4980AL (Keysight, USA) and a Keithley picoampere-meter (model 6485) (Keithley Instruments, Cleveland. OH, USA) at room temperature. Moreover, cyclic deformations with simultaneous testing of the electrical parameters of the material or structures were carried out.

Home-made installations were used for applying tensile and compressive strains to the tested films. The strain was estimated with the use of well-known equation $\varepsilon = (d + t)/2r$, where d is the thickness of the flexible substrate (d = 245 mm), t is the film thickness, and r is the substrate bending radius. The film thickness t < 100 nm could be neglected in comparison with the substrate.

## 3. Results and Discussion

### 3.1. CVD Grown Graphene and Multilayer Graphene Films

The AFM images of the CVD-grown graphene films transferred onto the SiO$_2$/Si and, with the use of a laminator, to the polyethylene terephthalate (PET) substrates are given in Figure 1a,b. The Raman spectrum for graphene on SiO$_2$/Si (Figure 1c) demonstrates its high quality. A standard set of the peaks (defect-related D, C-C-related G-peak and the second harmonic 2D-peak) was observed in the spectra. The resistivity as a function of the fluorination time for one structure on the SiO$_2$/Si substrates shows that a 9-min treatment in a weak solution of hydrofluoric acid is enough for the transition from the conductive to the insulating state. Generally, this transition corresponds to the fluorination degree F/C ~ 25%. The fluorination degree was estimated based on the X-ray photoemission data [27,28]. The insulating film, immediately after the resistivity enlargement, consists of graphene islands embedded into a fluorinated graphene matrix. It should be noted that the fluorination process starts at domain boundaries and introduces an additional surface corrugation [27]. For graphene on PET, the increase in the resistivity was not as strong as in the case of the rigid substrate: resistivity only had an increase of about 1.0–1.5 orders of magnitude.

The SEM and AFM surface images of the CVD grown multilayer graphene film transferred to SiO$_2$/Si substrates and, with the use of the laminator, to PET are demonstrated in Figure 2. The image analysis demonstrates that the grown MG consists of pyramid-like domains. A set of monolayer terraces is clearly visible in the AFM image measured in the regime of frictional forces (Figure 2c). The MG thickness in pyramid tops was about 8–10 nm.

The Raman spectra for the CVD-grown MG on the SiO$_2$/Si substrate are given in Figure 2d. Attention should be paid to the fact that the D peak intensity is relatively low, in spite of the pyramid-like domain structure of MG.

The dependences of resistivity as a function of fluorination time (treatment in the water solution of fluorine acid [27,28]) for the similar structures of MG on SiO$_2$/Si and PET substrates are demonstrated in Figure 3. One can see that, for the MG film on the SiO$_2$/Si substrate, the resistivity is strongly increased within a few minutes of fluorination. In the case of the flexible substrate, similar and longer treatments lead to a weak increase in the resistivity. In a few samples, we observed a strong increase in resistivity from 1–2 up to $10^3$ kOhm/sq under fluorination. Generally, an equally weak increase in the resistance was found for both graphene and MG (compare Figures 1d and 2) under the fluorination treatment on a flexible substrate.

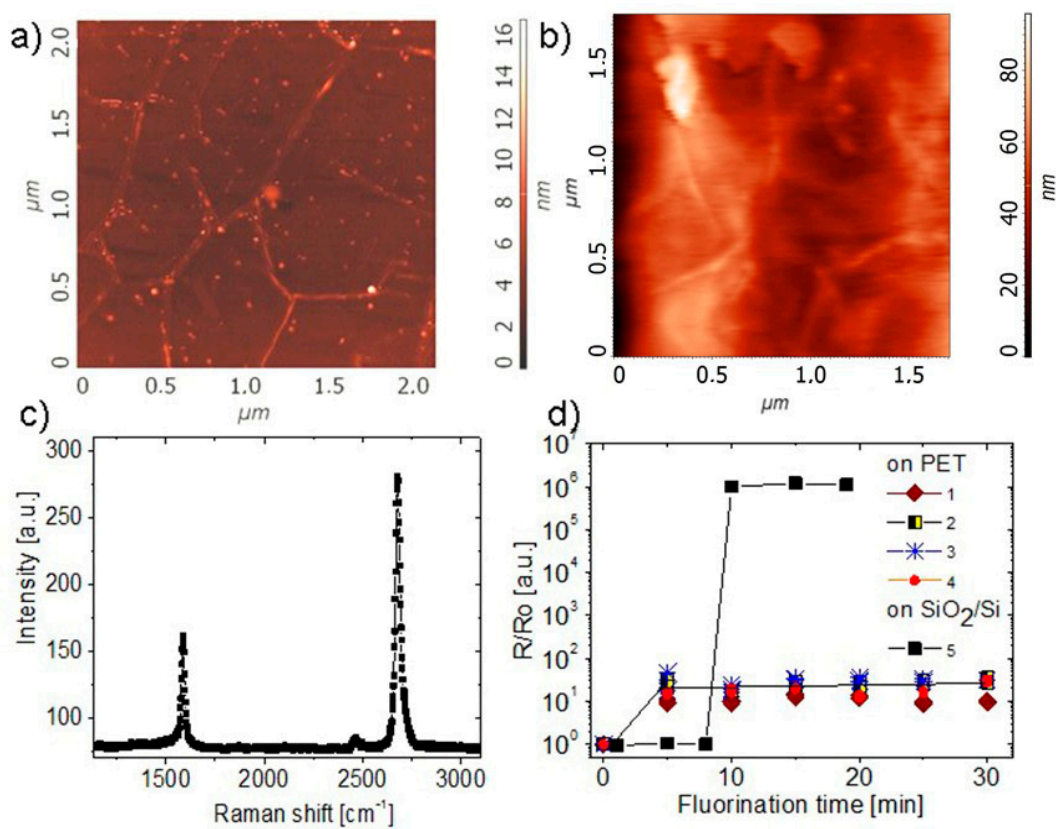

**Figure 1.** (**a**,**b**) Atomic force microscopy (AFM) images of CVD-grown graphene films transferred to the SiO$_2$/Si and PET substrates (**c**) The Raman spectra for graphene on SiO$_2$/Si substrate. (**d**) Resistivity as a function of fluorination time for few (1–4) structures on PET and one structure on SiO$_2$/Si substrates for comparison. R$_o$ for both graphene samples was ~900 Ohm/sq.

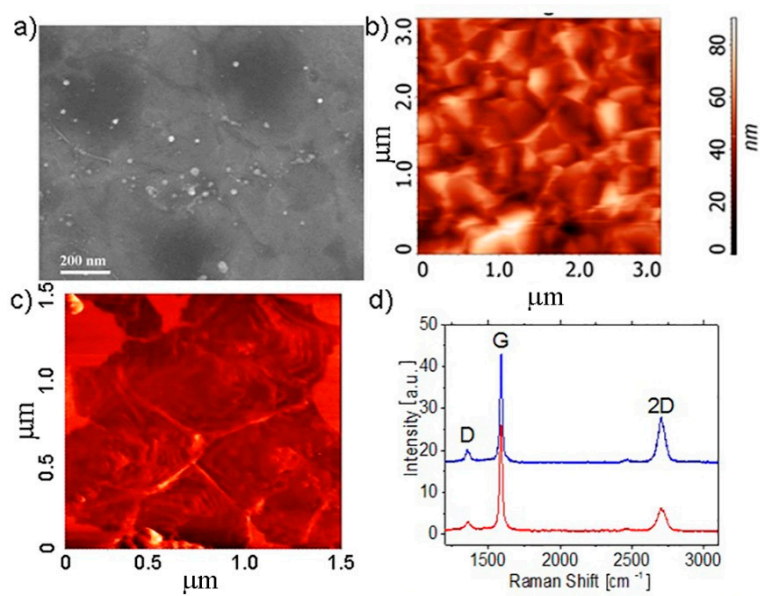

**Figure 2.** The scanning electron microscopy (SEM) (**a**) and AFM (**b**,**c**) images of CVD-grown multilayer graphene (MG) films transferred to the SiO2/Si (**c**) substrates and PET (**b**) with the use of a laminator. The AFM images were measured in the regime of the height (**b**) and the frictional forces (**c**). The MG thickness in pyramid tops was about 8-10 nm. (**d**) The Raman spectra for MG on SiO2/Si substrate measured in the two different points.

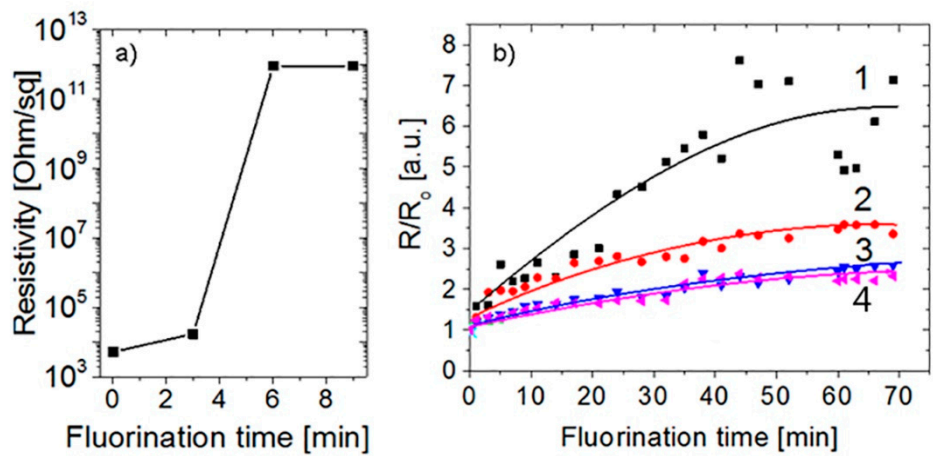

**Figure 3.** Changes in the resistivity under the fluorination process for the CVD-grown MG transferred onto (**a**) the SiO$_2$/Si and (**b**) PET substrates. In the case of (**b**), different curves correspond to the different tested structures.

The fluorination process is known to start with edge defects. The MG transferred on a flexible substrate turns over itself and the pyramid-like part of the film is pressed into the glue layer on PET. But even the top MG layer, which provides the main contribution to the resistivity value, weakly changes its resistance in this case. So, the fluorination process is weakened or even nearly suppressed in the case of graphene and MG on PET. We assumed that this effect is associated with the use of a laminator when the film is fixed on a substrate using glue. As we showed earlier, the fluorination of graphene films is accompanied by the graphene surface corrugation, since it is the graphene bending that removes the ban on the fluorination reaction [27]. In the case of glue, using this very process is suppressed.

The different changes in the resistance depended on the sample size. A relatively weak change in the resistivity given in Figure 3b was observed in the case when the sample was relatively small (few millimeters). In larger samples, non-glued areas provide a more pronounced increase in the resistance.

Changes in the resistivity for graphene under the tensile and compressive strain due to the bending of graphene on PET are presented in Figure 4a. Resistance keeps the constant value under the tensile deformations and, only after the compressive strain with a strong decrease in R, resistance starts to increase during the next cycle of the tensile deformation. A decrease in the resistance under the tensile deformation was most likely caused by a formation of the conductive paths through FG between the graphene islands. Fluorine atoms at a relatively low fluorination degree form an open fluorinated network with the defects at the domain boundaries in our polycrystalline graphene. Compression leads to the creation of additional conductive paths in this network. The increase in the resistance after a few bending cycles is generally found.

The MG films were also tested under tensile and compressive bending deformations. The results of repeated bending measurements for the pristine MG are given in Figure 4b. Generally, after one to two tuning cycles of the repeated measurements, a change in the film resistivity dependence versus bending strain is observed and a new type of characteristic becomes reproducible. The decrease in the MG length leads to a decrease in the scale of resistance change under bending. The results for the two fluorinated structures are given in Figure 4c,d. The first type of resistance dependences (Figure 4c) was observed for samples with low resistivity value (60–70 kOhm/sq, low fluorination degree). For the 11-min fluorinated sample, the resistivity dependence as a function of strain was practically the same as that before the fluorination. This fact is most likely associated with the suppression of the fluorination process for the sample well bonded to the adhesive layer. In the second case (Figure 4d), the sample resistivity was 10$^3$ kOhm/sq (high fluorination degree). In this sample, some changes in the resistance (25–50%) were observed at a strain of about zero; for a higher strain, the resistance was constant (with a relatively high scattering of the points at the tensile strain). Thus, it was found that the scale of resistivity changes under bending is dependent on the sample scale and fluorination degree. A higher fluoridation degree results in a weaker change in the resistance under bending.

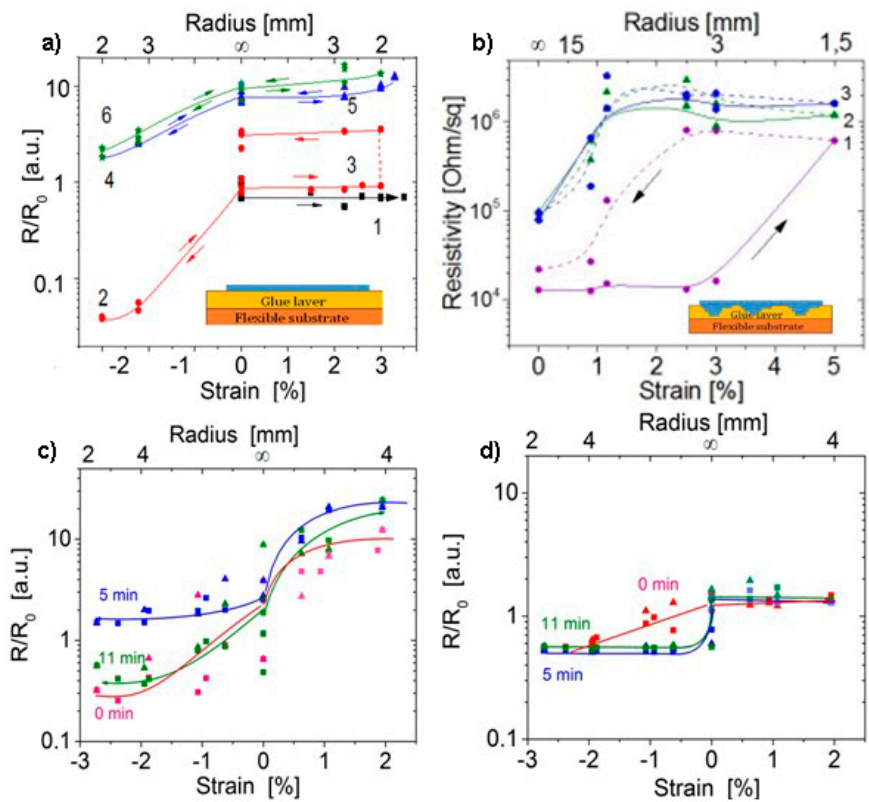

**Figure 4.** Changes in the resistivity under the tensile (+) and compressive (-) strain due to the bending of (**a**) fluorinated graphene (the fluorination time was 10 min) and (**b**–**d**) fluorinated MG on PET. Initial resistances were (**a**) $R_0$ = 2.0 kOhm, (**c**) $R_0$ = 67 kOhm/sq, (**d**) $R_0$ = $10^3$ kOhm/sq. (**a**,**b**) The sequence of deformations and measurements is indicated by numbers. The fluorination time is given in (**c**,**d**) at the curves as a parameter. Inserts in (**a**) and (**b**) presents a sketchy image of the tested structures.

### 3.2. Films Created from FG Suspension

The images of a series of the structures prepared on flexible polyimide (PI), PET, and paper substrates are demonstrated in Figure 5. On the above substrates, the structures containing Ag contacts shaped as two interpenetrating combs (Figure 5a) with three to four interdigital fingers were created by means of 2D printed technologies. Then, fluorinated graphene suspensions with the fluorination degree of F/C = 20–25% were applied onto the structures by drops. The thickness of these films was 30–50 nm. The high transmission of FG films does not allow one to see the presence of FG thin film on the surface of contacts and substrates in the optical images. The bending of the structures in the radius range of 10–1 mm for the used substrates corresponds to a tensile strain of 1.7–10%.

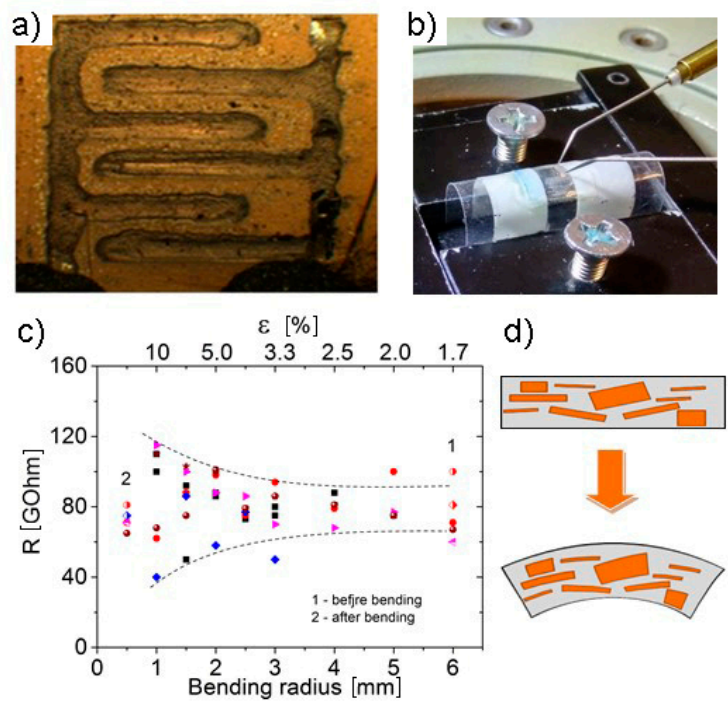

**Figure 5.** The optical image of (**a**) the structure created from the partially fluorinated graphene suspension with a thickness of about 30 nm on PET with Ag contacts and (**b**) shaped as two interdigital fingers. (**c**) Film resistance versus bending radius and strain at repeated measurements. Points 1 and 2 correspond to resistance before and after the bending. (**d**) Schematic representation of a film formed from flakes before and during bending.

The current–voltage characteristics of the structures with FG films measured at different bending radii were proved to be linear. From these characteristics, the resistance R of the structure was calculated to obtain the resistance dependence on the substrate bending radius (Figure 5c). Notable changes in the resistance were observed at bending radii less than 2.5 mm (strains greater than 4%). Repeated current measurements showed a relatively high data scattering both in the initial resistance of the structures and, essentially, in the resistance values at the minimum bending radii of 1 mm. The scale of resistance changes $\Delta R/R_0$ was in the range of 14–28%, with a different sign at 10 tensile strains (bending radius was 1 mm). These values correspond to typical ones for the changes in the graphene resistivity [19,29]. The scattering of the initial resistance values at large bending radii or without bending is most likely connected with a high resistivity of fluorinated films and was found to be within ±15%. In addition, a comparison of the resistance values before and after a measurement cycle showed that the resistance of the structure was almost not different from the initial value. Multiple bendings up to the 10% strain did not result in the film breakage; they did not cause notable changes in their properties after several tens of 100 deformation cycles with a strain value of ±0.15%. So, there is no degradation of the film properties under the action of cycles of relatively weak tensile or compressive strains.

### 3.3. FG films in Two-Layer Memristor Structures

It was found that the FG structures printed on polyvinyl alcohol films demonstrate a stable bipolar resistive switching effect with the ON/OFF current ratio of four to five orders of magnitude [30]. The FG fluorination degree was ~20–25%, and it corresponds to the case when FG flakes have their graphene islands embedded into the fluorinated matrix. The crossbar structures Ag/FG/PVA/Ag (active layer fluorinated graphene/polyvinyl alcohol) were made by printing on the surface of the flexible substrate (polyimide). The printed two-terminal device structures, the scheme, and images of the FG/PVA surface are shown in Figure 6. Due to the FG/ PVA film transparency, the printed active layer is not seen in the figure and is given schematically for one of the structures. The area of working structures was $60 \times 60$ $\mu m^2$, the PVA thickness was about 100 nm and the FG-layer thickness, according to the AFM measurements, for the similar film on the Si substrate, was ~4 nm. If one takes into account the fact that the porous PVA structure formed due to the dissolution of the PVA layer in water during the FG printing, the thickness of FG in crossbar structures was lower than 4 nm.

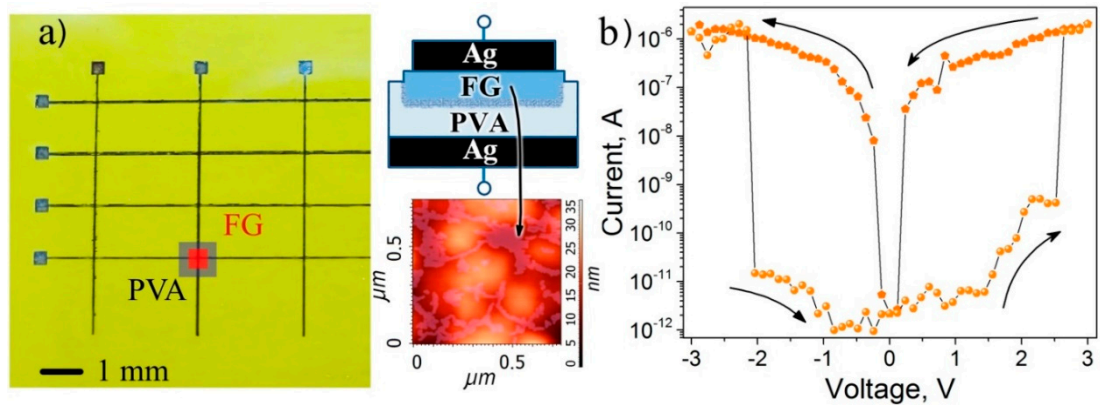

**Figure 6.** (**a**) The vertical crossbar Ag/ PFG/PVA /Ag structures on the polyimide (PI) substrate. The schematic cross-section of the printed structure and image of the active layer surface is shown in the insets. (**b**) The current–voltage characteristics for the printed crossbar structure measured before the deformation.

The effect of stability during mechanical deformations was investigated for printed crossbar structures under tensile bending conditions. The resistance switching before bending is given in Figure 6b. As shown in Figure 6, the structures retain their performance up to the 6.5% deformation, corresponding to the bending radius of 2 mm, and they restore it when the mechanical strain is removed.

The resistance switching observed at different deformation stages and the one after the strain removal are also given in Figure 7. The decrease in the ON/OFF current ratio observed with the increasing deformation to more than 6.5% manifests itself in a strong decrease in the ON current of three orders of magnitude. It should be noted here that the structures continue to switch at a strain of 6.5–8%, but with the ON/OFF current ratio of about two orders of magnitude. The conductivity in our crossbar structures is found on graphene islands in the FG matrix and traps at the FG/PVA interface. The analysis of the transport properties of these structures, in combination with charge spectroscopy, which was done in [30], demonstrates that graphene islands and traps at the FG interface with polyvinyl alcohol take part in resistive switching. As is demonstrated in Figures 6 and 7, the switching voltage was 2–3 V and did not notably change under strain. The decrease in ON current is most likely connected with an increase under the tensile strain in the distances between conductive agents in the films and the thickness of FG barriers between them.

The mechanism of the resistive switching is most likely related to the formation of localized states at the FG/PVA interface, in turn leading to the formation of new flow paths for the electric current [30]. Thus, on the one hand, PVA takes part in the formation of surface states, as it increases the ON/OFF current ratio. On the other hand, PVA plays the role of a developed elastic and porous substrate. The strong increase in current is associated with percolation transition in active layer i.e., with the appearance of the conductivity paths.

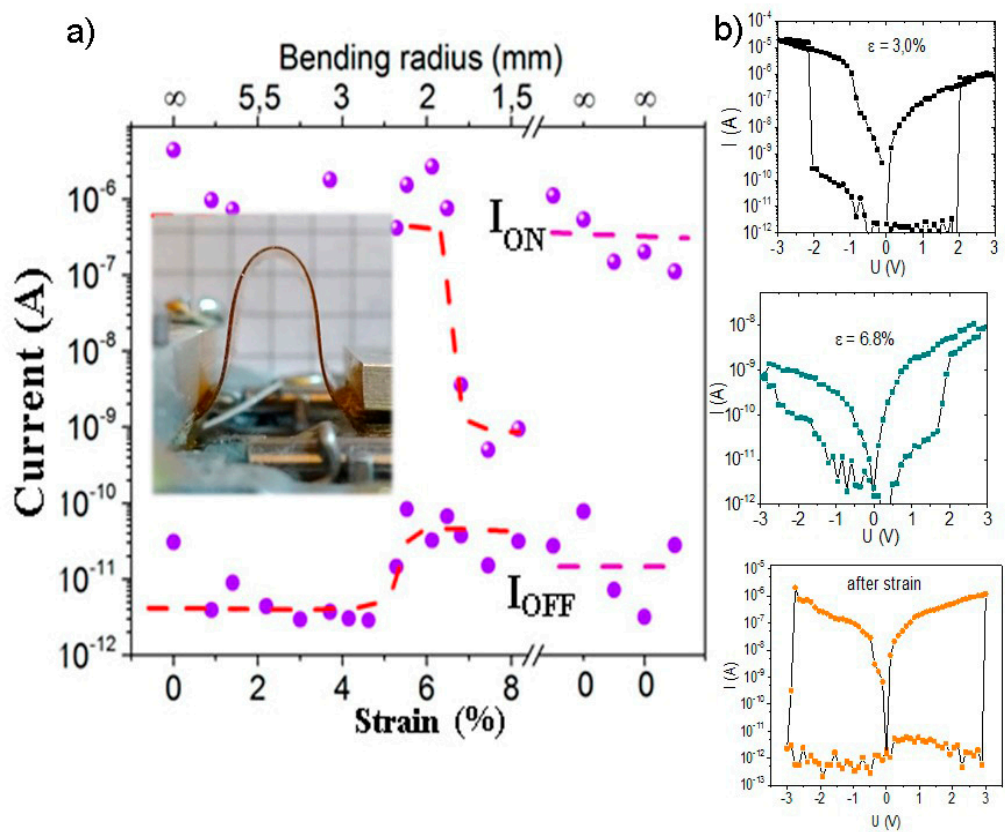

**Figure 7.** (**a**) The dependence of the current in the low ON and high OFF resistance states on the tensile strain and the bending radius of the crossbar structure. After x-axis breaking, few values of the $I_{on}$ and $I_{off}$ currents for the structure after removal of the strain are given. The inset shows a photo image of the deformed structure; (**b**) The current–voltage characteristics for the printed crossbar structure measured at the different tensile strain and after deformation.

## 4. Discussion

The first finding demonstrated in the present study is the weakening or even the suppression of the graphene and multigraphene fluorination process in the case of the strong connection between graphene and substrate. In this case, only one-side fluorination is expected. The variation in the resistance magnification is most likely determined by the quality of the connection with the substrate (see the explanatory illustration in Figure 8). The stronger the connection with the substrate, the lower the fluorination degree and the lower the film resistance after fluorination.

In the case of graphene, it is a well-known result when its resistivity is practically constant up to the tensile strain of 12% [20]. For the films from fluorinated graphene, there is no information about their mechanical properties. However, it is believed that the FG behavior will be similar to graphene or graphene oxide, and a good flexibility is expected.

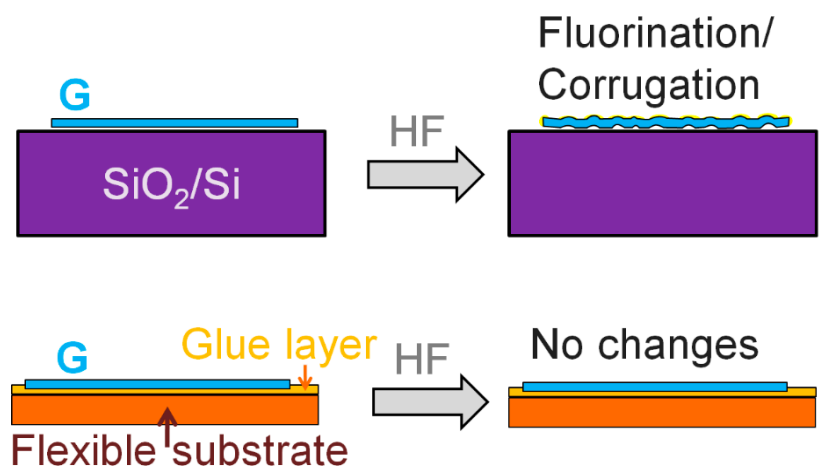

**Figure 8.** Schematic representation of the differences in the graphene fluorination process on the rigid and flexible substrates.

Generally, the structural details of the tested films determine the response during bending. Our graphene and MG films are polycrystalline. It is the defects at the domain boundaries and the plane edges, and their passivation or functionalization by fluorine atoms, that will be responsible for the film behavior at bending conditions. In our experiments, we used graphene (defects only at the domain boundaries), MG films (defects both at the domain boundaries and the plane edges) and the films from a suspension (defects only at the plane edge). It was experimentally observed that even a weak fluorination passivates all types of defect. However, the different reactions relating to bending were found for films with various structures.

Following from Figure 4a,d and Figure 6, the tensile bending of graphene or MG films with a relatively high fluorination degree leads to practically constant resistance values, similar to graphene. A strong decrease in the resistance under the compressive strain was observed for both graphene and MG films. The film fluorination results in a shift of the range when the resistance is reduced to lower strain values. The network of defects at the domain boundaries is, in this case, the "bottleneck" controlling the current in the film. The fluorine passivation of defects facilitates the formation of additional current paths in the film at compression conditions.

In the case of the films created from the suspension, it is important to mention here that these films consist of a large number of flakes, and the basal planes of flakes are located not only horizontally on the substrate, some of the flakes have different orientations. As a result of the film bending, the flakes are subjected to different types of deformation (symmetric and asymmetrical stretching, shear and non-uniform deformations). According to the theoretical study, the asymmetrical stretching, shear and non-uniform deformations lead to the strongest bending effects connected with the bandgap opening [20–22,31–33]. The resulting bending effect was found to vary under repeated measurements on such a mesoscopic system. Even small changes in the tested structure fixation in the holder result in a variable reaction to the bending. In general, it was these films that, despite some variation in their resistance values, showed the best deformation stability. These are the films that were used to create the memristor crossbars.

Good operation of the memristor crossbar structures created from the FG suspension under the tensile deformation is ensured by the stability of these film properties found in the present study, although some scattering in the values of ON and OFF currents is observed in these structures. Nevertheless, the excellent performance of the memristors up to the 6% strain, and even up to the 8% strain without failure, and the complete recovery after the deformation removal show that such films and structures are promising for applications

## 5. Conclusions

A change in the resistivity of different structures containing the fluorinated graphene films and flakes with various fluorination degrees under tensile and compressive strain due to bending deformations were studied. The graphene and multilayer graphene films grown by means of the CVD method transferred onto a flexible substrate by laminating and then subjected to the fluorination demonstrate a relatively weak increase in the resistance due to a weak fluorination degree. The fluorination process is weakened or even nearly suppressed in the case of graphene and MG on a flexible substrate with the use of a laminator when the film is fixed on a substrate using glue. These films consist of graphene islands and the network of the fluorinated graphene formed at domain boundaries or defects. The resistivity of fluorinated CVD-grown graphene and multilayer graphene films strongly decreases under the compressive strain. This decrease in resistance under tensile deformation was most likely caused by the formation of the conductive paths through FG between graphene islands. Tensile deformation leads to a relatively weak but irreversible increase in the resistivity. For FG films printed from the suspension with a fluorination degree 20–25%, a stable resistivity, at least up to 2–3% tensile and compressive strain, was found. The scale of resistance changes ($\Delta R/R_0$) was found to be in the range of 14–28%, with the various signs at the 10% tensile strain (bending radius 1 mm). In the case of the structures with the FG thin film printed on polyvinyl alcohol, a stable bipolar resistive switching was observed up to 6.5% of the tensile strain (bending radius was about 2 mm). A further increase in strain (6.5–8%) leads to a decrease in the ON/OFF current ratio from five down to two orders of magnitude. The current ratio decrease is connected with an increase, under the tensile strain, in the distances between conductive agents (graphene islands and traps at the interface with polyvinyl alcohol) within the active layer. The excellent performance of the memristors under deformation shows that FG films and structures created from suspension are especially promising for applications.

**Author Contributions:** Conceptualization, I.A.; Investigation, N.N., N.Z., I.K. and A.I.; Supervision, I.A.; Validation, N.N.; Visualization, A.I.; Writing—original draft, I.A.; Writing—review & editing, I.A. All authors have read and agreed to the published version of the manuscript.

**Acknowledgments:** The authors are thankful to D.Nikolaev and S.A.Smagulova from Ammosov North-Eastern Federal University (Yakutsk, Russia) for the provision of CVD-grown graphene.

**Conflicts of Interest:** The authors declare no conflict of interest.